\def\MPL #1 #2 #3 {Mod.~Phys.~Lett.~{\bf#1},\  #2 (#3)}
\def\NPB #1 #2 #3 {Nucl.~Phys.~{\bf#1},\  #2 (#3)}
\def\PLB #1 #2 #3 {Phys.~Lett.~{\bf#1},\  #2 (#3)}
\def\PR #1 #2 #3 {Phys.~Rep.~{\bf#1},\ #2 (#3)}
\def\PRD #1 #2 #3 {Phys.~Rev.~{\bf#1},\  #2 (#3)}
\def\PRL #1 #2 #3 {Phys.~Rev.~Lett.~{\bf#1},\  #2 (#3)}
\def\RMP #1 #2 #3 {Rev.~Mod.~Phys.~{\bf#1},\  #2 (#3)}
\def\ZP #1 #2 #3 {Z.~Phys.~{\bf#1},\  #2 (#3)}
\def\IJMP #1 #2 #3 {Int.~J.~Mod.~Phys.~{\bf#1},\  #2 (#3)}

\def\etc{{\it etc.}}
\def\dmgamgam{\Delta\mgamgam}
\def\ptmin{p_T^{\rm min}}
\def\mgamgam{M_{\gam\gam}}
\def\dmgamgam{\Delta\mgamgam}

\def\wstar{W^{\star}}

\def\zstar{Z^{\star}}

\def\br{B}

\def\rts{\sqrt s}

\def\h{h}

\def\gamh{\Gamma_{\h}^{\rm tot}}
\def\gamhl{\Gamma_{\hl}^{\rm tot}}
\def\gamhsm{\Gamma_{\hsm}^{\rm tot}}

\def\eg{{\it e.g.}}

\def\epem{e^+e^-}

\def\lplm{\ell^+\ell^-}

\def\lsim{\mathrel{\raise.3ex\hbox{$<$\kern-.75em\lower1ex\hbox{$\sim$}}}}
\def\gsim{\mathrel{\raise.3ex\hbox{$>$\kern-.75em\lower1ex\hbox{$\sim$}}}}
\def\@versim#1#2{\vcenter{\offinterlineskip
        \ialign{$\m@th#1\hfil##\hfil$\crcr#2\crcr\sim\crcr } }}
\def\zstar{Z^\star}
\def\wstar{W^\star}

\def\ie{{\it i.e.}}

\def\gam{\gamma}

\def\anti{\overline}

\def\fbi{~{\rm fb}^{-1}}

\def\gev{\,{\rm GeV}}

\def\hsm{h_{SM}}
\def\mhsm{m_{\hsm}}
\def\hl{h^0}

\def\mhl{m_{\hl}}

\def\h{h}

\documentstyle[preprint,aps]{revtex}    

\def\ie{{\it i.e.}}

\def\9{\phantom 0}      
\renewcommand\linebreak{\unskip\break} 
\begin{document}
\input psfig.sty
\newlength{\captsize} \let\captsize=\small 
\newlength{\captwidth}                     

%
\font\fortssbx=cmssbx10 scaled \magstep2
\hbox to \hsize{
%
%
$\vcenter{
\hbox{\fortssbx University of California - Davis}
}$
\hfill
$\vcenter{
\hbox{\bf UCD-96-15} 
\hbox{July, 1996}
\hbox{Revised: April, 1997}
}$
}

%
\medskip
\begin{center}
\bf
Prospects for and Implications of
Measuring the Higgs to Photon-Photon Branching Ratio
at the Next Linear {\boldmath$\epem$} Collider
\\
\rm
\vskip1pc
{\bf John F. Gunion and Patrick C. Martin}\\
\medskip
{\em Davis Institute for High Energy Physics}\\
{\em Department of Physics, University of California, Davis, CA 95616}\\
\end{center}

\begin{abstract}
We evaluate the prospects for measuring $\br(\h\to\gam\gam)$
for a Standard-Model-like Higgs boson at the Next Linear $\epem$ Collider
in the $\epem\to\zstar\to Z\h$ and $\epem\to\nu_e\anti\nu_e\h$ production modes.
Relative merits of different machine energy/luminosity strategies
and different electromagnetic calorimeter designs are evaluated.
We emphasize the importance of measuring $\br(\h\to\gam\gam)$ in order to obtain
the total width of a light Higgs boson and thereby the $b\anti b$ partial
width that will be critical in discriminating between the SM Higgs
and the Higgs bosons of an extended model. 

\end{abstract}

\section{Introduction}

\indent\indent 
One of the most important tasks of a Next Linear $\epem$ Collider (NLC)
will be to detect and study Higgs boson(s). For any observed
Higgs boson, extraction of its fundamental couplings and total width
in a model-independent manner will be a primary goal.
Measurement of $\br(\h\to \gam\gam)$ turns out to be 
an absolutely necessary ingredient
in extracting the total width and $b\anti b$ coupling 
in the case of a light Higgs boson with mass $\lsim 130\gev$\footnote{For
$\mhsm\gsim 130\gev$, a 2nd technique based on $W\wstar$ decays
emerges \cite{snowmass96}.}
and couplings similar to those of the Standard Model (SM) Higgs, $\hsm$,
and therefore a total width that is too small to be directly observed.
The procedure for obtaining the total and $b\anti b$ partial
widths using $\br(\h\to\gam\gam)$ is the following:
\begin{itemize}
\item 
Determine $\br(\h\to b\anti b)$ in $\epem\to\zstar\to Z\h$ and $\epem\to \epem
\h$ ($ZZ$-fusion) from the ratios
$\br(\h\to b\anti b)=[\sigma(Z\h)\br(\h\to b\anti b)]/ \sigma(Z\h)$
(with $Z\to\lplm$, $\ell=e,\mu$) and $\br(\h\to b\anti
b)=[\sigma(\epem\h)\br(\h\to b\anti b)]/ \sigma(\epem\h)$, respectively.
For $L=200\fbi$ of data at $\rts=500\gev$, the error for $\br(\h\to b\anti
b)$ would be about $\pm 5\%$ \cite{snowmass96}.
\item
Measure at the associated $\gam\gam$ collider facility the
rate for $\gam\gam\to\h\to b\anti b$ (accuracy $\sim\pm8\%$ \cite{snowmass96}
for $L=50\fbi$) 
proportional to $\Gamma(\h\to\gam\gam)\br(\h\to b\anti b)$ and compute 
(accuracy $\sim \pm 13\%$)
$\Gamma(\h\to \gam\gam)=[\Gamma(\h\to\gam\gam)\br(\h\to b \anti b)] /
\br(\h\to b\anti b)$.
\item 
Measure $\br(\h\to\gam\gam)$ as described shortly, and then compute:
\begin{equation}
\gamh= {\Gamma(\h\to\gam\gam)\over \br(\h\to\gam\gam)}\,;~~{\rm and}~~
\Gamma(\h\to b\anti b)=\gamh \br(\h\to b\anti b) \;.
\label{stepiv}
\end{equation}
\end{itemize}
For a SM-like $\h$, 
measurement of $\br(\h\to\gam\gam)$ at the NLC will be challenging 
because of its small size (at best of order a few times $10^{-3}$
\cite{dpfreport}). One will measure
$[\sigma(\epem\to Z\h)\br(\h\to \gam\gam)]$,
$[\sigma(\epem\to \nu_e\anti\nu_e \h)\br(\h\to \gam\gam)]$ 
and $[\sigma(\epem\to \nu_e\anti\nu_e \h)\br(\h\to b\anti b)]$ 
(the latter two being $WW$-fusion processes)
and compute $\br(\h\to\gam\gam)$ via the $Z\h$ and $WW$-fusion ratios,
\begin{equation}
{[\sigma( Z\h)\br(\h\to
\gam\gam)]\over\sigma( Z\h)}\,~~{\rm and}~~
{[\sigma(\nu_e\anti\nu_e\h)\br(\h\to \gam\gam)]\br(\h\to
b\anti b)\over [\sigma(\nu_e\anti\nu_e\h)\br(\h\to b\anti b)]}\,,
\label{ways}
\end{equation}
respectively.
Errors in the above two $\br(\h\to\gam\gam)$ computations will be
dominated by the errors in the $\sigma\br(\h\to\gam\gam)$ measurements.
(The $\epem\h$ final state from $ZZ$-fusion provides a third
alternative, but does not yield competitive errors
because of a larger background.)
Which of the ratios in Eq.~(\ref{ways}) will yield the smallest errors 
for $\br(\h\to\gam\gam)$ is dependent upon many factors.
In this Letter, we assess the relative merits of the $Z\h$ and $WW$-fusion
modes as a function of Higgs boson mass, machine energy, electromagnetic
calorimeter resolution and luminosity/upgrade strategies.

The importance of a direct determination of $\gamh$ and
$\Gamma(\h\to b\anti b)$ is due to the ambiguities associated 
with measuring only $\br(\h\to b\anti b)$. Consider, for example,
the light $\hl$ of the minimal supersymmetric model (MSSM). Model
parameter choices are easily found such that $\Gamma(\hl\to b\anti b)$
is much larger than predicted for the $\hsm$ \cite{dpfreport},
but $\br(\hl\to b\anti b)$ is only slightly larger
than expected due to the fact that the numerator, $\Gamma(\hl\to b\anti
b)$, and denominator, $\gamhl$, are both increased by similar amounts.
Extra (supersymmetric particle) 
decay modes could even enhance $\gamhl$ further, 
and $\br(\hl\to b\anti b)$ could be smaller than the 
SM prediction despite the fact that $\Gamma(\hl\to b\anti b)$ is enhanced.
Equation~(\ref{stepiv}) shows that the
ability to detect deviations of $\gamh$ and $\Gamma(\h\to b\anti b)$ from
SM expectations depends critically on the error
in $\br(\h\to\gam\gam)$, which
is very likely to be the dominant source of uncertainty.
Of course, dramatic deviations of $\br(\h\to\gam\gam)$ from SM expectations
are also a possibility, even if the $\h$ is very SM-like
in its couplings to the SM particles.  Large effects can be caused by 
new particles (fourth generation, supersymmetric, \etc) 
in the one-loop graphs responsible for the $\h\to\gam\gam$ coupling.
Regardless of the size of the deviations from SM predictions,
determining $\br(\h\to\gam\gam)$ will be vital to
understanding the nature of the Higgs boson and will
provide an important probe of, or limits on,
new physics that may lie beyond the SM.

\section{Procedures}

\indent\indent
We consider SM Higgs masses in the range $70- 150\gev$;
$\br(\hsm\to\gam\gam)$ in units of $10^{-3}$ is
0.75, 1.0, 1.4, 1.8, 2.2, 2.6, 2.6, 2.2, 1.6 
as $\mhsm$ ranges from 70 to 150 GeV in steps of 10 GeV.
In computing signals and backgrounds, we use exact matrix elements.
To define $Z\gam\gam$ vs. $\nu_e\anti\nu_e \gam\gam$ events, we employ
the recoil mass, $M_{X}=\sqrt{(p_{e^+}+p_{e^-}-p_{\gam_1}-p_{\gam_2})^2}$.
We define $Z\gam\gam$ events as $X\gam\gam$ events for which $M_X$ is 
within the interval $[80,100]$ (GeV). In this way, 
we can use all $Z$ decay modes while ensuring that the only significant
background is that from $Z\gam\gam$ non-Higgs diagrams.
(Interference between signal and background $Z\gam\gam$ diagrams is small.)
The $M_X$ cut also implies that for $X=\nu_e\anti\nu_e$ the signal
is almost entirely from $\zstar\to Z\hsm$
(interference with the $WW$-fusion diagram being small).
Conversely, we define $\nu_e\anti\nu_e\gam\gam$ events as $X\gam\gam$ events 
($X=\nu_\ell\anti\nu_\ell$, $\ell=e,\mu,\tau$) such that $M_X\geq
130\gev$. This effectively leaves only the $WW$-fusion signal contribution
and non-$Z$-pole background diagrams; interference is again small.

In both the $Z\gam\gam$ and $\nu_e\anti\nu_e\gam\gam$ modes,
our goal will be to minimize the $\sigma\br(\h\to\gam\gam)$ error,
defined as $\sqrt{S+B}/S$, where $S$ ($B$) is the number
of Higgs signal (background) events. The first important choice is $\rts$.
For the $ Z\gam\gam$ channel, the optimal $\rts$ values
are given by $\rts_{\rm opt}(\mhsm)\sim 89\gev+1.25\mhsm$
(always close to the peak in the $Z\hsm$ cross section
and $\leq 300\gev$ for $\mhsm\leq 150\gev$).
For the $\nu_e\anti\nu_e\gam\gam$ mode, the smallest errors
are achieved when $\rts$ is as large as possible.
We give results for $\rts=500\gev$, at which $\rts$ 
the $Z\gam\gam$ channel also remains useful.
Next are the kinematical cuts.
Because of the small signal rates,
these cuts must be chosen to reduce the background as much
as possible while retaining a large fraction of the Higgs signal events.
Keeping in mind the fact that, as a function of $\mgamgam$,
the Higgs resonance sits on a slowly varying background, 
a very crucial cut is to accept only events in a small mode-, $\mhsm$- and 
detector-resolution-dependent (see later discussion) interval of $\mgamgam$ 
centered on $\mhsm$,\footnote{The
Higgs mass will be very precisely measured at the NLC.
The background level under the peak will be very precisely normalized
using measurements with $\mgamgam$ away from $\mhsm$.}
with width chosen so as to minimize $\sqrt{S+B}/S$.
Additional one-dimensional and two-dimensional kinematic cuts for minimizing
the error were extensively investigated. 
\begin{itemize}
\item
For the $Z\hsm$ mode, the best cuts we found are the following:
\begin{equation}
 p_T^{\gam_{1,2}}\geq {\mhsm\over 4}\,,\quad
p^{\gam_1}_T+p^{\gam_2}_T\geq \ptmin(\mhsm)\,,
\label{cutsi}
\end{equation}
where $p_T^{\gam_{1,2}}$ are the transverse momenta
of the two photons in the $\epem$ center-of-mass. 
(By convention,  $E_{\gam_1}\geq E_{\gam_2}$.)
Within the statistics of our Monte Carlo study,
the optimal $\ptmin$ values at $\rts=\rts_{\rm opt}$ ($\rts=500\gev$)
are given by $\ptmin(\mhsm)\sim 0.9\mhsm - 10\gev$ ($\ptmin(\mhsm)\sim 200\gev$);
for such $\ptmin$, the photon rapidities are always within 
$|y_{\gam_1}|\leq 1.2$ and $|y_{\gam_2}|\leq 1.6$.
\item
In the $\nu_e\anti\nu_e\hsm$ mode,
the smallest error was achieved using the following cuts:
\begin{eqnarray}
&|y_{\gam_1}|\leq 2.5 \,,\quad |y_{\gam_2}|\leq 2.5\,,& \nonumber\\
& p_{T}^{\gam_{1,2}}\geq p_{T}^{\gam_{1,2}~{\rm min}}(\mhsm)\,,\quad
p_{T}^{\gam_1}+p_{T}^{\gam_2} \geq \ptmin(\mhsm)\,,&
\label{cutsii} \\
&p_{T}^{\rm vis}=
\sqrt{(p_x^{\gam_1}+p_x^{\gam_2})^2+(p_y^{\gam_1}+p_y^{\gam_2})^2}
\geq 10 \gev\,.&\nonumber
\end{eqnarray}
Within our Monte Carlo statistics,
the optimal numerical choices (at $\rts=500\gev$)
as a function of $\mhsm$ are described by:
$p_T^{\gam_1~{\rm min}}(\mhsm)\sim 0.16\mhsm+20\gev$,
$p_T^{\gam_2~{\rm min}}(\mhsm)\sim 0.18\mhsm+1\gev$,
 and $\ptmin(\mhsm)\sim 0.5\mhsm+35\gev$.
The $p_{T}^{\rm vis}$ cut is needed to eliminate reducible backgrounds
due to events such as $e^+e^- \to e^+e^- \gam\gam $
where the $e^+$ and $e^-$ are lost down the beam pipe
leaving the signature of $\gam\gam$ plus missing energy \cite{ptvs}.
\end{itemize}
We note that after the cuts of Eq.~(\ref{cutsi}) or Eq.~(\ref{cutsii}),
the photons have substantially different energies,
especially in the $WW$-fusion case.

Four different electromagnetic calorimeter resolutions
are considered:
(I) resolution like that of the CMS lead tungstate crystal \cite{CMS}
with $\Delta E/E=2\%/\sqrt E\oplus 0.5\%\oplus20\%/E$;
(II) resolution of $\Delta E/E=10\%/\sqrt E\oplus 1\%$; 
(III) resolution of $\Delta E/E=12\%/\sqrt E\oplus 0.5\%$; and
(IV) resolution of $\Delta E/E=15\%/\sqrt E\oplus 1\%$.
Cases II and III are at the `optimistic' end of current NLC detector
designs \cite{NLCdetector}. 
Case IV is the current design specification for the 
JLC-1 detector \cite{JLCdetector}. For each resolution case
and choice of $\mhsm$, we determined the $\dmgamgam$ value which
minimizes $\sqrt{S+B}/S$ in the $Z\gam\gam$ and $\nu_e\anti\nu_e\gam\gam$ modes.
The optimal $\dmgamgam$ values for the $Z\h$ mode at $\rts=\rts_{\rm opt}$
and the $WW$-fusion mode at $\rts=500\gev$
are the same within Monte Carlo errors: 
$\dmgamgam({\rm I,II,III,IV})({\rm GeV})\sim (0.015,0.035,0.035,0.045)\mhsm$.
For $Z\hsm$ production at $\rts=500\gev$, $\sqrt{S+B}/S$ is minimized for
$\dmgamgam({\rm I,II,III,IV})\sim (0.015,0.03,0.03,0.04)\mhsm$.

The optimal $\dmgamgam$, $p_T^{\rm min}$, and $p_{T}^{\gam_{1,2}~{\rm min}}$
values specified above are `soft';  changes in the $p_T$ cuts
by $\pm 5\gev$ or in $\dmgamgam/\mhsm$ by $\pm 0.005$ lead to $\leq 0.01$
change in $\sqrt{S+B}/S$.

\section{Results and Discussion}

\begin{figure}[htb]
\leavevmode
\begin{center}
\centerline{\psfig{file=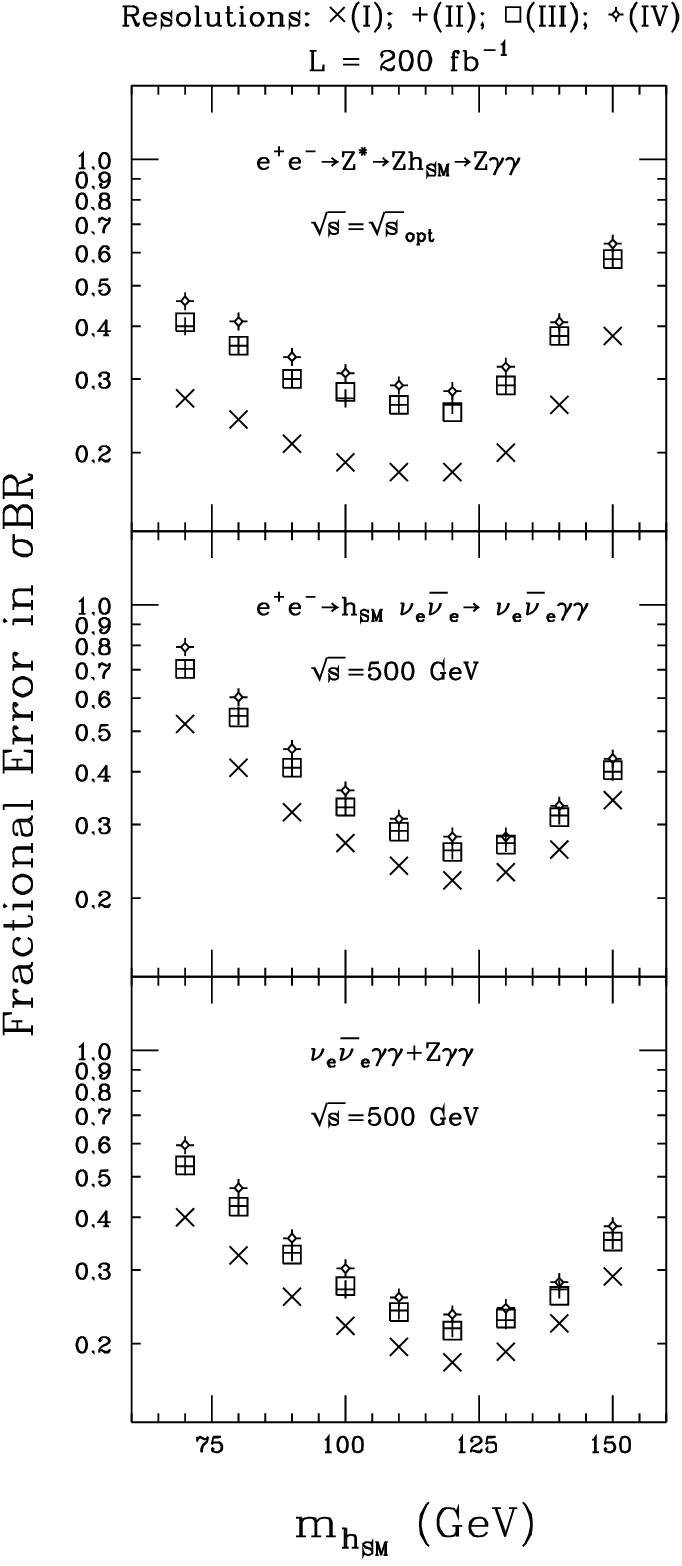,width=2.7in}}
\end{center}
\caption{The fractional error in the measurement of
$\sigma(\nu_e \bar\nu_e \hsm)\br(\hsm\to\gam\gam)$ (at $\protect\rts=500\gev$)
and $\sigma(Z\hsm)\br(\hsm\to\gam\gam)$ 
(at $\protect\rts=\protect\rts_{\rm opt}$)
as a function of $\mhsm$ assuming $L=200\fbi$.
Also shown is the fractional $\sigma\br(\hsm\to\gam\gam)$
error obtained by combining $Z\hsm$ and $\nu_e\anti\nu_e\hsm$
channels for $L=200\fbi$ at $\protect\rts=500\gev$.
Results for the four electromagnetic calorimeter 
resolutions described in the text are given.}
\label{fig1}
\end{figure}

The first two windows of Figure~\ref{fig1} show 
the statistical errors, $\sqrt{S+B}/S$,
for measuring $\sigma\br(\hsm\to\gam\gam)$ in the $\zstar\to Z\hsm$
and $\nu_e\anti\nu_e\hsm$ ($WW$-fusion) measurement modes
as a function of $\mhsm$. We assume
four years of $L=50\fbi/{\rm yr}$ running, \ie\ $L=200\fbi$,
at $\rts=\rts_{\rm opt}$ ($\rts=500\gev$) in the $Z\hsm$ 
($WW$-fusion) cases, respectively. Comparing,
we find that in resolution cases II-IV the $Z\hsm$ ($WW$-fusion)
measurement mode yields smaller errors for $70\lsim\mhsm\lsim 120\gev$
($130\lsim\mhsm\lsim 150\gev$).
In resolution case I, the $Z\hsm$ mode error is the
smaller for masses up to $130\gev$.
As a function of $\mhsm$,
the smallest errors are obtained for 
$100\gev\lsim \mhsm\lsim 130\gev$.\footnote{In
the MSSM the light Higgs has $\mhl\lsim 130\gev$.} For calorimeter
resolutions II or III, the errors range from
$\pm 25\%$ to $\pm 29\%$ for the ($\rts=\rts_{\rm opt}$) 
$Z\hsm$ measurement and from 
$\pm 26\%$ to $\pm 33\%$ for the ($\rts=500\gev$) $WW$-fusion measurement.

In the third window of Fig.~\ref{fig1} we plot
the error obtained by combining the $WW$-fusion and $Z\hsm$ mode
$\sigma \br(\hsm\to\gam\gam)$ 
statistics for $L=200\fbi$ accumulated at $\rts=500\gev$. 
\footnote{We do not discuss the reverse situation, 
since the $WW$-fusion rate at $\rts_{\rm opt}$ is always
$\lsim 1/5$ of that for $Z\hsm$.}
This is close to the error for $\br(\hsm\to\gam\gam)$ obtained
by combining the two ratios in Eq.~(\ref{ways})
given that errors for the other inputs are much smaller
than the $\sigma\br(\hsm\to\gam\gam)$ errors. Although the $Z\hsm$ mode
error at $\rts=500\gev$ is always larger than the $WW$-fusion mode error,
including the $Z\hsm$ measurement 
substantially improves the net $\br(\hsm\to\gam\gam)$ error relative to 
that obtained using $WW$-fusion alone,
especially at low $\mhsm$. For $100\gev\lsim\mhsm\lsim 130\gev$,
the net error ranges from $\pm 23\%$ to $\pm 27\%$.

Although observation of a clear Higgs signal
in the $\gam\gam$ invariant mass distribution
is not an absolute requirement (given that we will have observed the $\hsm$
in other channels and will have determined its mass very accurately)
it would be helpful in case there is significant systematic uncertainty in
measuring the $\gam\gam$ invariant mass.  It is vital to be
certain that the $\dmgamgam$ interval is centered on the mass region where
the Higgs signal is present. Taking
$\rts=\rts_{\rm opt}$ ($\rts=500\gev$) for the $Z\hsm$
($WW$-fusion) mode and $L=200\fbi$, we find $S/\sqrt B\geq 3$ in the following 
(resolution-dependent) regions:
\begin{eqnarray}
{\rm case~I:} &~~~& 70\leq\mhsm\leq150\gev~(Z\hsm),~~80\leq\mhsm\leq150\
\gev (WW{\rm-fusion})\,,\nonumber
\nonumber\\
{\rm cases~II/III:} &~~~& 80\leq\mhsm\leq140\gev~(Z\hsm),~~90\leq\mhsm\leq150\
\gev (WW{\rm-fusion})\label{3sig}\\
{\rm case~IV:} &~~~& 90\leq\mhsm\leq130\gev~(Z\hsm),~~100\leq\mhsm\leq150\
\gev (WW{\rm-fusion})\,,\nonumber
\end{eqnarray}

\section{Final Remarks and Conclusions}

\indent\indent
We have studied the prospects for 
measuring $\sigma\br(\h\to\gam\gam)$ for a SM-like Higgs boson,
with $70\leq\mhsm\leq150\gev$, at the NLC.
The measurements will be challenging but of great importance.
We have compared results for two different production/measurement modes:
$\zstar\to Z\h$ and $WW$-fusion. Errors for the $WW$-fusion
channel are minimized at full machine energy, $\rts=500\gev$.
Errors in the $Z\h$ channel are minimized if the machine energy
is tuned to the ($\leq 300\gev$) $\rts=\rts_{\rm opt}$ value which maximizes 
the $Z\h$ event rate.  The net error for $\br(\hsm\to\gam\gam)$
is approximately given by combining the $WW$ and $Z\h$ channel
$\sigma\br$ errors, since errors for other quantities entering
the ratios of Eq.~(\ref{ways}) are small. At $\rts=500\gev$, 
the error obtained using only the
$WW$-fusion channel measurement is significantly decreased by including
the $Z\h$ channel measurement.
At $\rts=\rts_{\rm opt}$, the $WW$-fusion channel can be neglected and the 
net error is essentially just that for the $Z\h$ channel.
At any $\rts$ and in either channel, the better the electromagnetic 
calorimeter resolution, the smaller the error in $\br(\hsm\to\gam\gam)$.
For $100\leq\mhsm\leq 130\gev$,
where $\br(\hsm\to\gam\gam)$ is largest (a mass range that 
is also highly preferred for the light SM-like $\hl$ of the MSSM),
and $L=200\fbi$, the net error assuming an excellent CMS-style calorimeter 
(resolution case I) falls in the ranges
$\sim\pm 18\%$ to $\sim\pm20\%$ at $\rts=\rts_{\rm opt}$
and $\sim\pm 18\%$ to $\sim\pm22\%$ at $\rts=500\gev$.
For $L=200\fbi$ and a calorimeter at the
optimistic end of current plans for the NLC detector (cases II and III),
the $100\leq\mhsm\leq 130\gev$ net error falls in the ranges
$\sim\pm 25\%$ to $\sim\pm 29\%$ at $\rts=\rts_{\rm opt}$ and
$\sim\pm 22\%$ to $\sim\pm 27\%$ at $\rts=500\gev$.

If the NLC is first operated at $\rts=500\gev$,
either because a Higgs boson has not been detected previously
or because other physics (\eg\ production of supersymmetric particles)
is deemed more important, data will be accumulated
with whatever calorimeter is part of the initial detector
and a corresponding measurement of $\br(\hsm\to\gam\gam)$ will result. 
The desirability of stopping data collection to upgrade the
calorimeter and/or reconfigure the interaction region for full
luminosity at the $Z\hsm$ cross section maximum must be carefully 
evaluated.\footnote{It is best to continue to run at $\rts=500\gev$
if the interaction region is not reconfigured for full luminosity
at the lower $Z\hsm$-channel $\rts_{\rm opt}$.}
Using the $L=200\fbi$ errors of Fig.~\ref{fig1}, we find that 
it is {\it not} advantageous to reconfigure for
$\rts=\rts_{\rm opt}$ if $\mhsm\gsim 100\gev$. The value
of a calorimeter upgrade is also marginal for such $\mhsm$. To illustrate,
suppose the initial calorimeter has resolution II or III.
For $\mhsm=120\gev$, upgrading the calorimeter from II/III to I,
and then accumulating a 2nd $L=200\fbi$ at $\rts=500\gev$ after doing so, 
would yield a net $\br(\hsm\to\gam\gam)$ error 
of $\pm 14\%$, as compared to $\sim\pm 15.5\%$ 
if no changes are made and a total
of $L=400\fbi$ is accumulated by simply running twice as long.
For $\mhsm=150\gev$, upgrading the resolution
would yield (after the 2nd $L=200\fbi$ run at $\rts=500\gev$) 
$\sim\pm 22\%$ error vs. $\sim\pm 25\%$ if no calorimeter change is made.
However, for small $\mhsm$, reconfiguration 
and high resolution calorimetery both become quite valuable 
at the NLC. For example, if $\mhsm=70\gev$ (roughly the current LEP I/II 
limit), a 2nd $L=200\fbi$ run with full ${\cal L}$ at $\rts=\rts_{\rm opt}$
and upgrade to resolution I would yield $\sim\pm24\%$ error 
vs. $\sim\pm 38\%$ after a 2nd $L=200\fbi$ run with no changes. For
$\mhsm=70\gev$,
running from the beginning for $L=400\fbi$ at the $\sigma(Z\hsm)$ peak 
(as possible at full luminosity if $\mhsm$ is known from LHC data) 
with resolution I yields error of $\sim\pm 19\%$.

In evaluating different options/strategies, 
it is necessary to keep in mind that LHC data may allow a rather competitive 
error for $\br(\hsm\to\gam\gam)$ \cite{snowmass96}. One combines
the $L=600\fbi$ (for ATLAS and CMS combined) LHC
measurement of $\br(\hsm\to\gam\gam)/\br(\hsm\to b\anti b)$ 
with the $L=200\fbi$, $\rts=500\gev$ NLC measurement of $\br(\hsm\to b\anti b)$
to obtain a value for $\br(\hsm\to\gam\gam)$ with error
$\sim \pm 16\%$ for $80\leq \mhsm\leq 130\gev$, rising
to $\sim\pm 25\%$ for $\mhsm\sim 140\gev$. If we combine this
$\br(\hsm\to\gam\gam)$ error
with the net error for the ($Z\hsm$ plus $WW$-fusion mode, $\rts=500\gev$,
$L=200\fbi$, resolution II/III)
direct $\br(\hsm\to\gam\gam)$ measurement at the NLC,
the overall error for $\br(\hsm\to\gam\gam)$ will be:
\begin{center}
\begin{tabular}{cccccccc}
$\mhsm$~(GeV) & 80 & 100 & 110 & 120 & 130 & 140 & 150 \\
Error         & $\pm 15\%$ & $\pm 14\%$, & $\pm 13\%$ & $\pm 13\%$ & $\pm 13\%$
              & $\pm 18\%$ & $\pm 35\%$ \\
\end{tabular}
\end{center}
For most Higgs masses, there would be little to gain from excellent (case I)
resolution. For example, at $\mhsm\sim 120\gev$,
the above $\sim \pm 13\%$ found assuming NLC resolution cases II/III 
would only improve to $\sim \pm 12\%$ for NLC resolution case I.
For $\mhsm\sim 80\gev$, the NLC $\hsm\to\gam\gam$ decay determination of
$\br(\hsm\to \gam\gam)$ will only be of value if $L=400\fbi$ with
calorimeter resolution I can be accumulated 
by the time $L=300\fbi$ per detector is accumulated at the LHC.
Finally, if determining $\gamhsm$, and thence $\Gamma(\hsm\to b\anti b)$,
is the dominant motivation for measuring $\br(\hsm\to\gam\gam)$,
then it is important to note that for $\mhsm\gsim 130\gev$
$\gamhsm$ is better determined
using the $\hsm\to W\wstar$ techniques discussed in Ref.~\cite{snowmass96}.
For such $\mhsm$, this fact and the small gain in $\br(\hsm\to\gam\gam)$
error (especially if LHC data is available)
argue against considering a calorimeter upgrade.

\section{Acknowledgements}

This work was supported in part by Department of Energy under
grant No. DE-FG03-91ER40674
and by the Davis Institute for High Energy Physics. 
We would like to thank T. Barklow, J. Brau, P. Rowson, R. Van Kooten
and L. Poggioli for helpful communications.

\clearpage

\begin{thebibliography}{99}

\bibitem{snowmass96} J.F. Gunion, L. Poggioli and R. Van Kooten, 
{\it Higgs Boson Discovery and Properties}, hep-ph/9703330,
to appear in {\it Proceedings of the 1996 DPF/DPB Summer Study
on ``New Directions in High Energy Physics'' (Snowmass, 96)}, June 25 - July
12, 1996, Snowmass, Colorado.

\bibitem{dpfreport} See J.F. Gunion, A. Stange, and S. Willenbrock,
{\it Weakly-Coupled Higgs Bosons}, preprint UCD-95-28 (1995),
to be published in {\it Electroweak Physics and Beyond the Standard Model}, 
World Scientific Publishing Co.,
eds. T. Barklow, S. Dawson, H. Haber, and J. Siegrist, a references therein.

\bibitem{ptvs} C.-H. Chen, M.Drees, and J.F. Gunion,
\PRL 76 2002 1996 .

\bibitem{CMS} CMS Technical Proposal, CERN/LHCC 94-38 (1994).

\bibitem{NLCdetector} 
{\it Physics and Technology of the Next Linear Collider:
a Report Submitted to Snowmass 1996}, BNL 52-502, p.~25;
see also the simulation homepage:
http:// nlc.physics.upenn.edu /nlc /software /software.html.

\bibitem{JLCdetector}
JLC-I, JLC Group, KEK Report 92-16, as summarized by K. Fujii, {\it Proceedings
of the 2nd International Workshop on ``Physics and Experiments
with Linear $\epem$ Colliders''}, eds. F. Harris, S. Olsen,
S. Pakvasa and X. Tata, Waikoloa, HI (1993), World
Scientific Publishing, p.~782.



\end{thebibliography}
\end{document}